\documentclass{article}

\usepackage{PRIMEarxiv}

\usepackage[utf8]{inputenc} % allow utf-8 input
\usepackage[T1]{fontenc}    % use 8-bit T1 fonts
\usepackage{hyperref}       % hyperlinks
\usepackage{url}            % simple URL typesetting
\usepackage{booktabs}       % professional-quality tables
\usepackage{amsfonts}       % blackboard math symbols
\usepackage{nicefrac}       % compact symbols for 1/2, etc.
\usepackage{microtype}      % microtypography
\usepackage{lipsum}
\usepackage{fancyhdr}       % header
\usepackage{graphicx}       % graphics
\usepackage{amsmath}
\usepackage{makecell}
\usepackage{subcaption}
\usepackage{enumitem}
\usepackage{listings}  % Load the package
\usepackage{xcolor}     % For syntax highlighting
\usepackage{textcomp}
\usepackage{gensymb}

\graphicspath{{media/}}     % organize your images and other figures under media/ folder

\title{CTorch: PyTorch-Compatible GPU-Accelerated Auto-Differentiable Projector Toolbox for Computed Tomography}

\author{
  Xiao~Jiang \\
  Department of Biomedical Engineering \\
  Johns Hopkins University \\
  Baltimore, MD, United States\\
  \texttt{xjiang43@jhu.edu} \\
   \And
  Grace~J.~Gang \\
  Department of Radiology \\
  University of Pennsylvania \\
  Philadelphia, PA, United States\\
  \texttt{grace.j.gang@pennmedicine.upenn.edu} \\
   \And
  J.~Webster~Stayman \\
  Department of Biomedical Engineering \\
  Johns Hopkins University \\
  Baltimore, MD, United States\\
  \texttt{web.stayman@jhu.edu} \\
}

\begin{document}
\maketitle

\begin{abstract}
This work introduces CTorch, a PyTorch-compatible, GPU-accelerated, and auto-differentiable projector toolbox designed to handle various CT geometries with configurable projector algorithms. CTorch provides flexible scanner geometry definition, supporting 2D fan-beam, 3D circular cone-beam, and 3D non-circular cone-beam geometries. Each geometry allows view-specific definitions to accommodate variations during scanning. Both flat- and curved-detector models may be specified to accommodate various clinical devices. CTorch implements four projector algorithms: voxel-driven, ray-driven, distance-driven (DD), and separable footprint (SF), allowing users to balance accuracy and computational efficiency based on their needs. All the projectors are primarily built using CUDA C for GPU acceleration, then compiled as Python-callable functions, and wrapped as PyTorch network module. This design allows direct use of PyTorch tensors, enabling seamless integration into PyTorch’s auto-differentiation framework. These features make CTorch an flexible and efficient tool for CT imaging research, with potential applications in accurate CT simulations, efficient iterative reconstruction, and advanced deep-learning-based CT reconstruction. 

CTorch is available at \hyperlink{https://github.com/JHU-AIAI-Shared/AIAI-CTorch}{https://github.com/JHU-AIAI-Shared/AIAI-CTorch} for non-profit, non-commercial use. Please email \hyperlink{web.stayman@jhu.edu}{web.stayman@jhu.edu} for access.

\end{abstract}

\section{Introduction}
Over the past half-decade, Computed Tomography (CT) technology has experienced significant advancements, leading to widespread applications in both medical imaging and industrial imaging fields\cite{withers2021x, de2014industrial}. As the name suggests, computing plays a crucial role in CT imaging, particularly in system simulation\cite{wu2022xcist} and image reconstruction\cite{hsieh2013recent}. Both processes rely on the fundamental operations of forward projection (where object attenuation values are integrated over lines in space to form projection data) and backprojection\cite{long20103d} (where projection data values are assigned along lines in a volume). These operations are computationally intensive and often constitute the efficiency bottleneck in CT imaging algorithms.

While incorporating projectors for projection simulation is relatively straightforward, their role in image reconstruction is more complex. Analytical reconstruction methods\cite{zou2004exact} primarily depend on backprojection, whereas model-based iterative reconstruction (MBIR)\cite{tilley2017penalized} requires both forward and backprojection to compute the derivatives of the reconstruction objective function and form iterative updates. Although traditional end-to-end deep learning reconstruction (DLR)\cite{chen2017low} does not explicitly require projectors, advanced model-based deep learning reconstruction integrates these operations to enforce data consistency\cite{jiang2024strategies} or perform domain transforms\cite{lin2019dudonet}. Moreover, Both MBIR and DLR typically require a gradient-based optimizer for either image reconstruction or network training, where the gradient computation can be cumbersome, especially when backpropagating the projector gradients through deep neural networks.

The auto-differentiation mechanism in PyTorch\cite{paszke2017automatic} significantly simplifies gradient computation by enabling automatic gradient backpropagation through a computational graph built from predefined differentiable operators. While directly implementing a projector in PyTorch ensures differentiability, this approach is inefficient due to the challenge of fully vectorizing all operations within the projection algorithm. A more effective solution is to implement low-level functions in CUDA C, compile them into Python-callable APIs, and wrap them as differentiable PyTorch operators. This strategy enables direct access to PyTorch tensor data and seamless integration with PyTorch’s built-in functions, ensuring efficient GPU-based computations.

Several Python tomography toolboxes with GPU acceleration have been developed, some of which are integrated with PyTorch \cite{van2015astra,biguri2016tigre,ronchetti2020torchradon,chen2023soul,kim2023differentiable}. Ideally, a projector toolbox should be capable of handling arbitrary CT geometries encountered in real-world scenarios, supporting both 2D and 3D projections. While flat detector geoemtries are widely implemented, the ability to model curved detectors is important due to their widespread use in diagnostic CT scanners. In flat-panel cone-beam CT (FP-CBCT), scanning trajectories may be non-circular\cite{gang2022universal}, and factors such as patient motion and gantry jitter often necessitate view-specific geometry descriptions to ensure accurate simulation and reconstruction. Furthermore, various well-established projector algorithms exist\cite{peters1981algorithms,siddon1985fast,long20103d,de2004distance}, allowing users to balance accuracy and efficiency based on specific needs. Unfortunately, most available packages offer limited flexibility in choosing projection algorithms, restricting customization for different applications.

To address these challenges, we introduce \textbf{CTorch}, a PyTorch-compatible, GPU-accelerated, and auto-differentiable projector toolbox for Computed Tomography. The key features of CTorch are:\vspace{-0.02in}
\begin{itemize}[leftmargin=0.5cm]
    \item \textbf{Flexible Geometry Definition}: Supports 2D fan-beam, 3D circular cone-beam, and 3D non-circular cone-beam geometries. Additionally, each geometry allows view-specific definitions to handle variations during scanning.
    \item \textbf{GPU-Accelerated Auto-Differentiable Forward/Back Projectors}: Directly accept variables as PyTorch tensors, enabling seamless integration into any PyTorch-based algorithm.
    \item \textbf{Multiple Projection Algorithms}: Implements four projector algorithms—voxel-driven, ray-driven, distance-driven (DD), and separable footprint (SF)—allowing users to select the most suitable method for their specific application.
    \item \textbf{Curved Detector Support}: Enables projection for 2D fan-beam and 3D circular cone-beam geometries, enhancing compatibility with clinical CT data.
    \item \textbf{Optimized Data Layout}: Follows PyTorch batched and channeled data format, facilitating efficient parallel processing of multiple samples and ensuring smooth integration with PyTorch-based pipelines.
\end{itemize}

\section{Methodology}
\label{sec:method}

\subsection{Toolbox Framework}
\begin{figure}[h]
    \centering
    \begin{minipage}{0.48\textwidth}
        \centering
        \includegraphics[width=\textwidth]{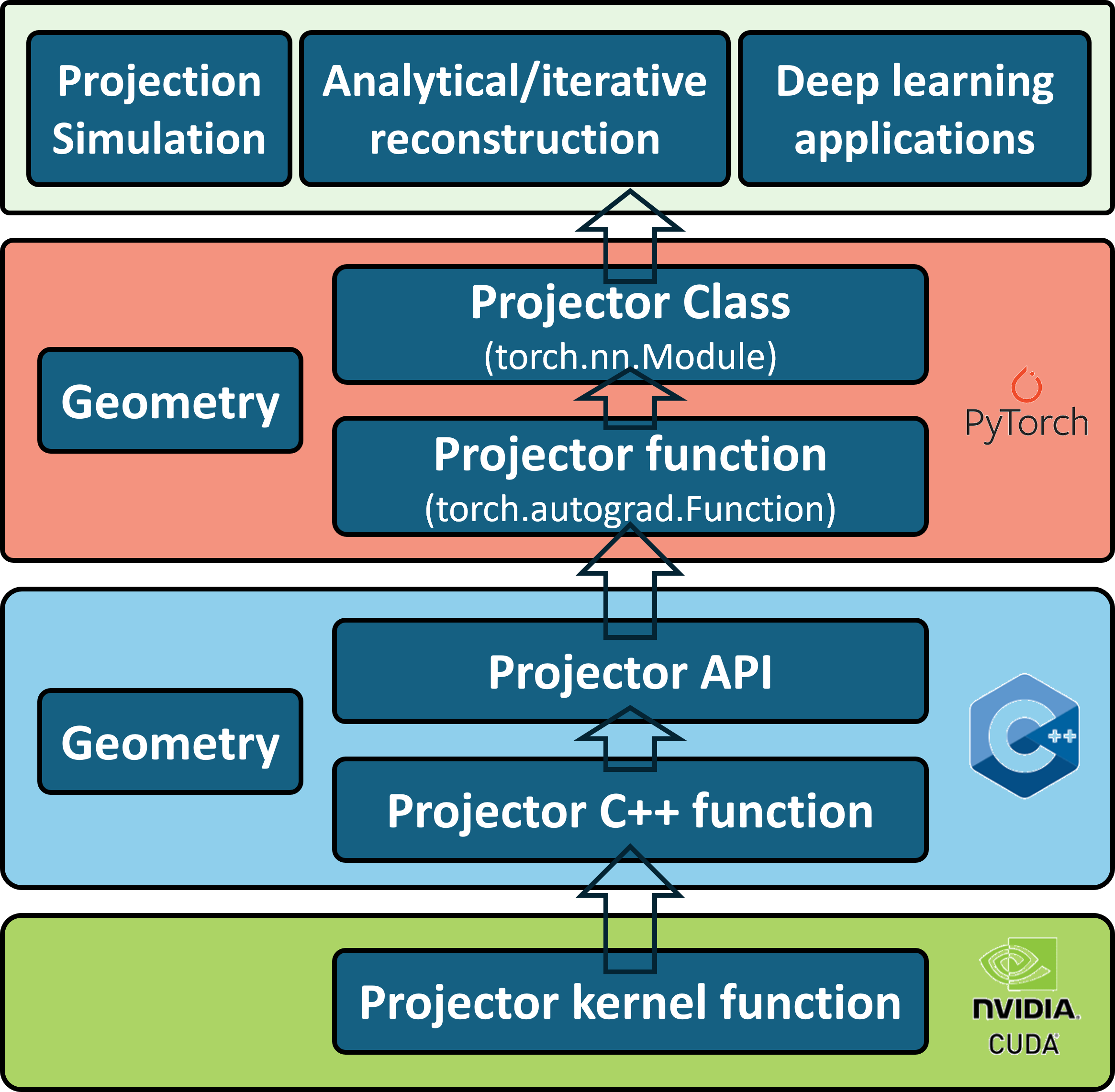}
        \caption{Overall framework of the CTorch toolbox. }
        \label{fig:framework}
    \end{minipage}\hfill
    \begin{minipage}{0.48\textwidth}
        \centering
        \includegraphics[width=\textwidth]{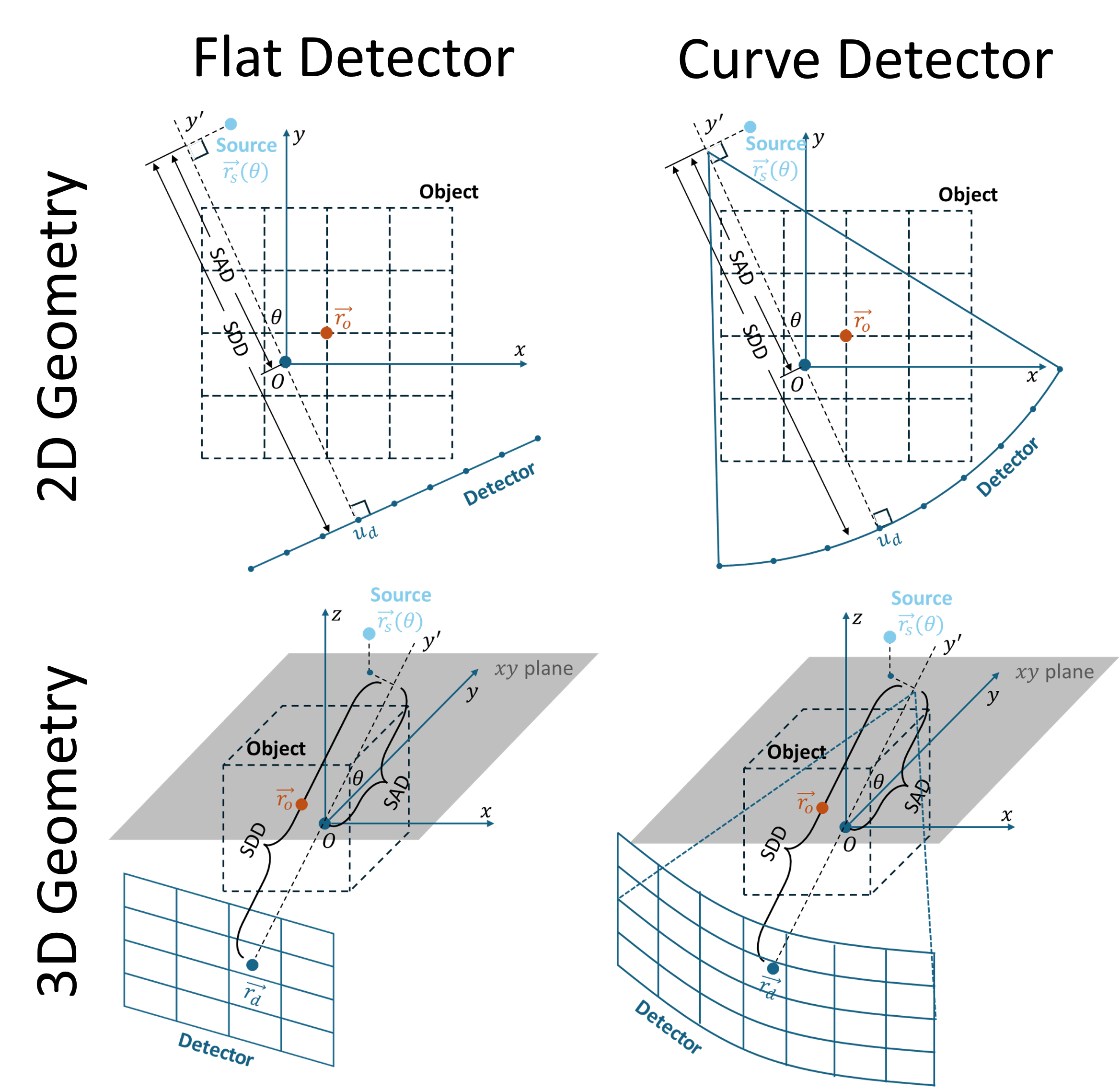}
        \caption{Schematic diagram of the 2D and 3D circular scan geometries and associated parameters.}
        \label{fig:geometry}
    \end{minipage}
\end{figure}

The overall framework of CTorch is summarized in Fig.~\ref{fig:framework}, where arrows indicate the direction of function calls. At the lowest level, CUDA kernel functions execute parallel computations on the GPU. C++ projector function manages kernel scheduling and constructs C++ geometry objects based on parameters passed from Python. This C++ projector function is then compiled into a Python-callable API using pybind11. Next, we define a custom PyTorch auto-differentiation function (inherited from torch.autograd.Function) by specifying both the forward and backward processes, both of which call the projector API. This ensures that the gradient of the projector is automatically computed when integrated into the computation graph. The PyTorch projector function is then incorporated into the forward function of a unified projector class (inherited from nn.Module), which provides high-level management of projection geometry, projector algorithm, and projection direction. Users only need to define the system geometry and instantiate Projector class, then incorporate the projector into their programs for various tomography tasks, including projection simulation and image reconstruction. To further facilitate these tasks, we have additionally implemented a discrete ramp filter and added ordered subsets method to the projector, along with example implementations of forward projection, filtered backprojection (FBP), model-based iterative reconstruction (MBIR), and Diffusion Posterior Sampling (DPS) reconstruction.

\subsection{Geometry Definition}
CTorch defines three types of geometries: 2D geometry, 3D circular scan geometry, and 3D non-circular scan geometry, covering a wide range of CT scanner configurations in real-world scenarios. The parameters for each geometry are summarized in Fig.~\ref{fig:geometry} and Table~\ref{tab:parameter}. The 2D Geometry is described by seven parameters: gantry rotation angle $\theta$, source-axis-distance(SAD), source-detector-distance(SDD), source lateral offset $x_{src}$ in world coordinates, object center position $(x_{ofst}, y_{ofst})$ in world coordinates, detector lateral offset $(u_{ofst})$ in world coordinates. The world coordinate origin is defined at the rotation center, with the y-axis perpendicular to the detector when $\theta=0$. All offset parameters are measured at $\theta=0$. Each geometry parameter can be specified as either a scalar or a vector: when a scalar is used, all views share the same geometry; when a vector is used, a view-specific geometry is defined for the scan. The 3D geometry extends the 2D geometry along the $z-$direction, adding three parameters: source vertical offset $z_{src}$, object vertical offset $z_{ofst}$, and detector vertical offset $v_{ofst}$. The 3D non-circular geometry further introduces two additional gantry rotation parameters ($\phi, \psi$), where $\theta, \phi,$ and $\psi$ define rotations around the $z-, x-, y-$axes, respectively. The overall rotation follows the order $y\rightarrow x \rightarrow z$. A detailed description of non-circular geometry can be found at \hyperlink{https://docs.openrtk.org/en/latest/documentation/docs/Geometry.html}{https://docs.openrtk.org/en/latest/documentation/docs/Geometry.html}.

All geometries in CTorch assume a single point-source model, meaning they follow a divergent beam geometry. While many existing packages provide a separate parallel beam geometry, the same effect can be achieved in CTorch by setting both SAD and SDD to sufficiently large values. For some unconventional CT scanner configurations, such as multiple sources and multiple detectors, users can define multiple projectors to emulate the system. It is also worth noting that a non-circular geometry reduces to a circular geometry with $\phi=\psi=0$, and the 3D geometry reduces to a 2D geometry with a single slice object/detector. However, instead of defining a universal geometry, we provide three geometries for two key reasons: 1. 2D geometry is widely used in deep learning-based CT research, and circular scans effectively describe many CT scanners in both simulations and real-world studies. Therefore, having three separate geometries simplifies configuration and improves usability. 2. According the properties of each geometry, we can develop targeted optimization strategies to maximize the projection computation speed. 
 
\begin{table}[h]
    \centering
    \renewcommand{\arraystretch}{1.5}
    \begin{tabular}{|c|c|c|c|}
        \hline
        & 2D Geometry & 3D Circular geometry & 3D Non circular \\
        \hline
        Gantry rotation & $\theta$(gantry) & $\theta$ & $\theta,$ $\phi$(offplane)$, \psi$(inplane)\\
        \hline
        Source-axis distance & $SAD$ & $SAD$ & $SAD$ \\
        \hline
        Source-detector distance & $SDD$ & $SDD$ & $SDD$ \\
        \hline
        Source Position & $\vec{r}_s(0) = (x_{src}, SAD)$ & $\vec{r}_s(0) = (x_{src}, SAD, z_{src})$ & $\vec{r}_s(0) = (x_{src}, SAD, z_{src})$ \\
        \hline
        Object Position & $\vec{r}_o = (x_{ofst}, y_{ofst})$ & $\vec{r}_o = (x_{ofst}, y_{ofst}, z_{ofst})$ & $\vec{r}_o = (x_{ofst}, y_{ofst}, z_{ofst})$ \\
        \hline
        Detector Position & $u_d = -u_{ofst}$ & $\vec{r}_d = (-u_{ofst}, -v_{ofst})$ & $\vec{r}_d = (-u_{ofst}, -v_{ofst})$ \\
        \hline
    \end{tabular}
    \caption{Parameter definition for different geometries.}
    \label{tab:parameter}
\end{table}

\subsection{Projector Algorithm}
\begin{figure}[h]
    \centering
    \includegraphics[width=0.8\linewidth]{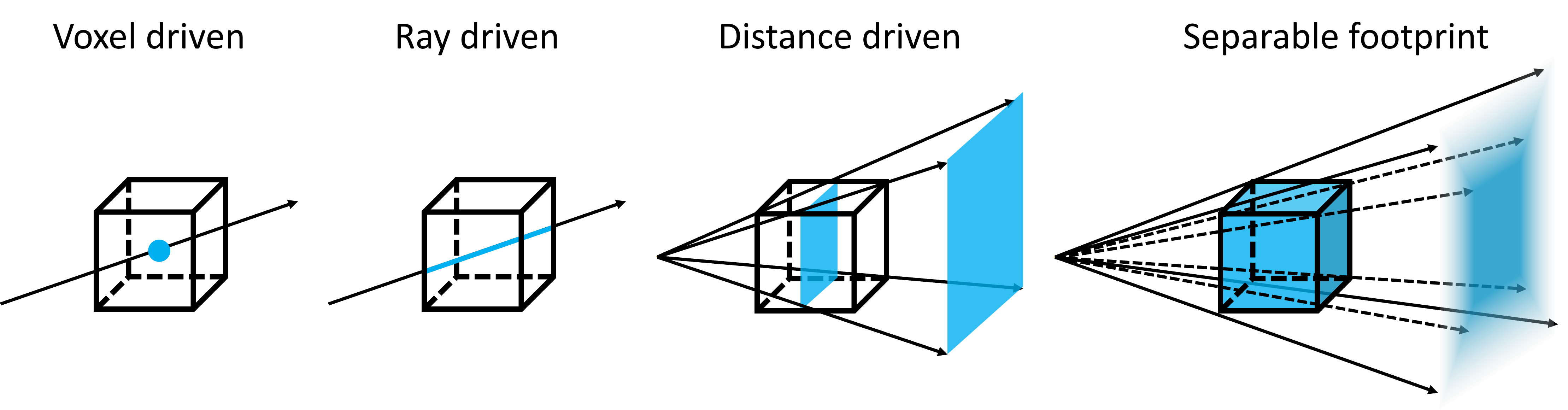}
    \caption{Illustration of forward projection of a single voxel using different projector algorithms.}
    \label{fig:projector}
\end{figure}
Given a discrete object volume $\textbf{x}$, the forward projection applies a linear operation to map the volume (performing line integrals) to a projection $\textbf{y}$:
\begin{equation}
    \textbf{y}=\textbf{Ax}
\end{equation}
where the \textbf{A} is determined by system geometry, and is commonly referred to as the system matrix. The backprojection is defined as the adjoint operation of the projection:
\begin{equation}
    \textbf{x}=\textbf{A}^T\textbf{y}
\end{equation}
which maps the projection back to the object volume. Although the system matrix is fully determined by the scanner geometry, it is typically too large to store, even with a sparse representation. Therefore, projector algorithms are generally designed to compute forward and backprojection on-the-fly. More importantly, forward projection and backprojection are gradients of each other since $\nabla_{\textbf{y}}\textbf{x}=\textbf{A}^T, \nabla_{\textbf{x}}\textbf{y}=\textbf{A}$, which means that a gradient computation can be performed simply by switching between forward projection and backprojection. This presumes the implementation of so-called "matched" projector and back-projector pairs.

CTorch provides four distinct projector algorithms, i.e., voxel driven\cite{peters1981algorithms}, ray driven\cite{siddon1985fast}, distance driven (DD)\cite{de2004distance}, and separable footprint (SF)\cite{long20103d}. As illustrated in Fig.~\ref{fig:projector}, voxel-driven, ray-driven, DD, and SF projector consider a point-, line-, face-, volume-based voxel model, respectively. Each projector algorithm and its strict adjoint(gradient) are implemented in CTorch. In terms of performance, ray-driven and voxel-driven methods offer fast computation, whereas distance-driven and separable footprint methods provide higher accuracy. Ray-driven, distance-driven, and separable footprint algorithms are mostly used in forward projection, while voxel-driven, distance-driven, and separable footprint algorithms are more suitable for backprojection. Additionally, the backprojection used in FBP reconstruction differs slightly from the general backprojection by incorporating a depth-dependent weighting\cite{kak2001principles}. To ensure high reconstruction accuracy, CTorch introduces two projection modes: "proj" mode, which follows standard forward and backprojection, and "recon" mode, which applies a depth-dependent weighting specifically for FBP reconstruction. The projectors for non-circular geometries are currently only implemented for flat-panel detectors.

\newpage
\subsection{Example}
\label{sec:example}
This section describes the basic usage of CTorch through a MBIR example. More examples can be found in the example folder in the toolbox.

\begin{lstlisting}
import torch
import CTorch.utils.geometry as geometry
from CTorch.projector.projector_interface import Projector

# geometry definition
geom = geometry.Geom2D(nx=256,ny=256,dx=0.5,dy=0.5,nu=384,du=0.5,nView=600,viewAngles=np.arange(0,-2*np.pi,-2*np.pi/600),detType="flat",
    SAD=[500],SDD=[1000],xOfst=[0.0],yOfst[0.0],uOfst=[0.0],xSrc=[0.0])    

# projector definition, forward projection with SF projector and "proj" mode
A = Projector(geom,'proj','SF','forward')

# create 256x256 object with a channel size of 3 and a batch size of 2 
image = torch.randn([2,3,256,256]).cuda().float()

# Forward projection sino = A img
sino = A(img)

# MBIR using gradient descent
num_iter, lr = 1000, 1.0 # number of iterations and step size
x = torch.zeros_like(img, requires_grad=True).cuda() # reconstruction

for i in range(num_iter):
    # compute least square loss
    loss = torch.nn.functional.mse_loss(A(x), sino)

    # automatic gradient computation
    grad = torch.autograd.grad(loss, [x], retain_graph=False)[0] 

    # gradient descent
    x.data -= lr * grad
\end{lstlisting}

\section{Performance}
\label{sec:benchmark}

\subsection{Projector and Reconstruction Accuracy}
\begin{figure}[h]
    \centering
    \begin{minipage}{0.48\textwidth}
        \centering
        \includegraphics[width=\textwidth]{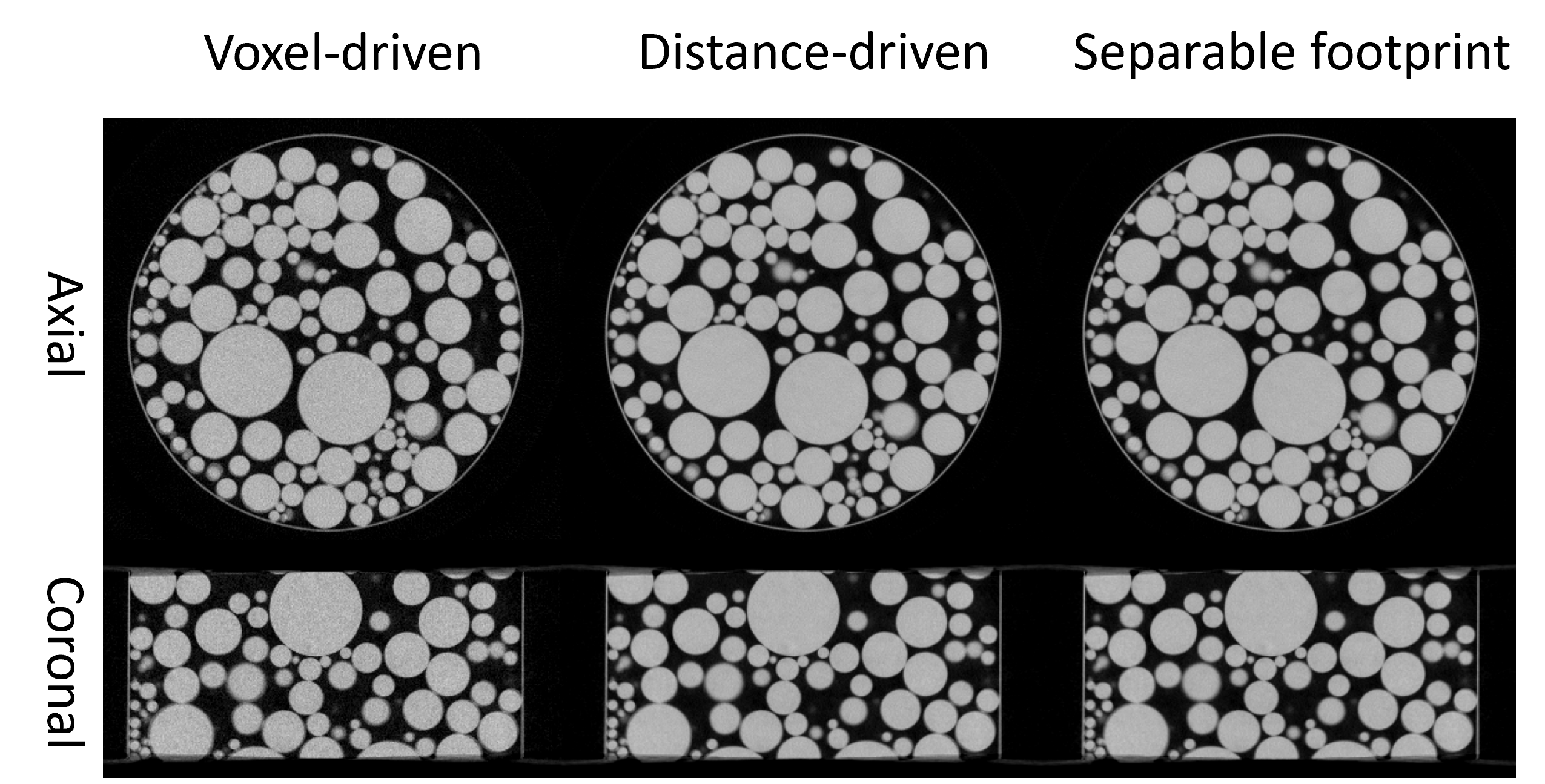}
        \caption{FBP reconstruction using the top-layer projections from a dual-layer CBCT system wiht different projector algorithms.}
        \label{fig:first}
    \end{minipage}\hfill
    \begin{minipage}{0.48\textwidth}
        \centering
        \includegraphics[width=\textwidth]{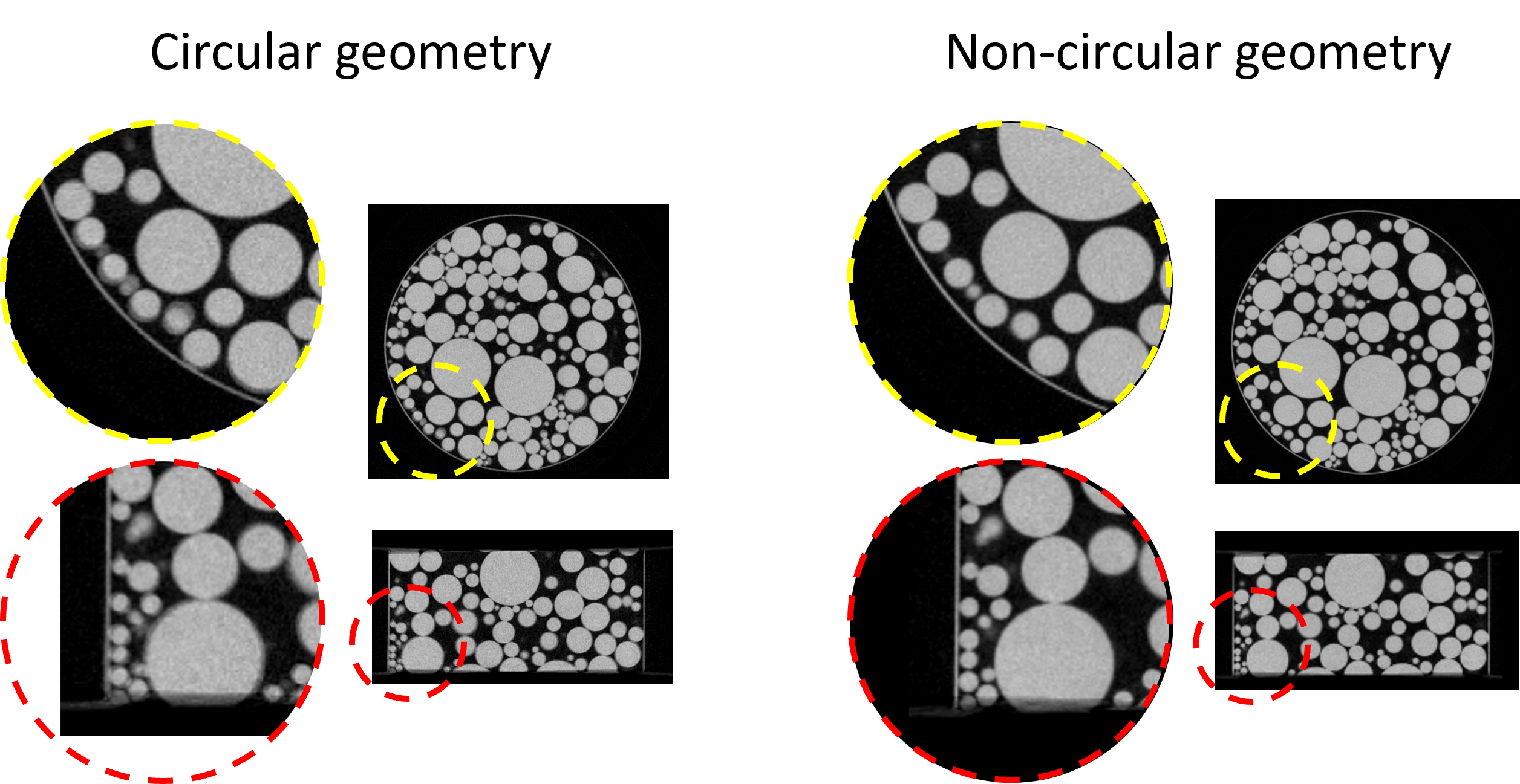}
        \caption{FBP reconstruction using the bottom-layer projections from a dual-layer CBCT system using different geometry definition. The non-circular geometry add a $0.4\degree$ in-plane rotation, effectively mitigating the edge blurring in the recon with circular geometry.}
        \label{fig:second}
    \end{minipage}
\end{figure}

We validated the CTorch projector for both circular and non-circular geometries using physical x-ray bench data. A plastic ball phantom was scanned on a laboratory benchtop CBCT system equipped with a dual-layer flat-panel detector\cite{CareView560DE}. The two-layer projections were reconstructed separately using the FBP algorithm under a 3D circular geometry assumption. Fig.~\ref{fig:first} presents the reconstruction results for the top layer using different projectors. Each projector accurately depicts the ball shape with sharp edges, demonstrating the high geometric accuracy of the CTorch circular-scan projector. Fig.~\ref{fig:second} shows the bottom-layer results. When the projections are backprojected under the circular geometry assumption, geometric distortions are noticeable, particularly around the ball edges. These distortions may arise from misalignment of the detector pixel grids between the two layers. However, by shifting to a non-circular geometry model that incorporates a $0.4\degree$ in-plane detector rotation, the distortions are substantially reduced. This validates the accuracy of the CTorch non-circular projector and highlights the importance of flexible geometry descriptions for processing nonideal physical projection data.

\begin{figure}[h]
    \centering
    \begin{minipage}{0.48\textwidth}
        \centering
        \includegraphics[width=\textwidth]{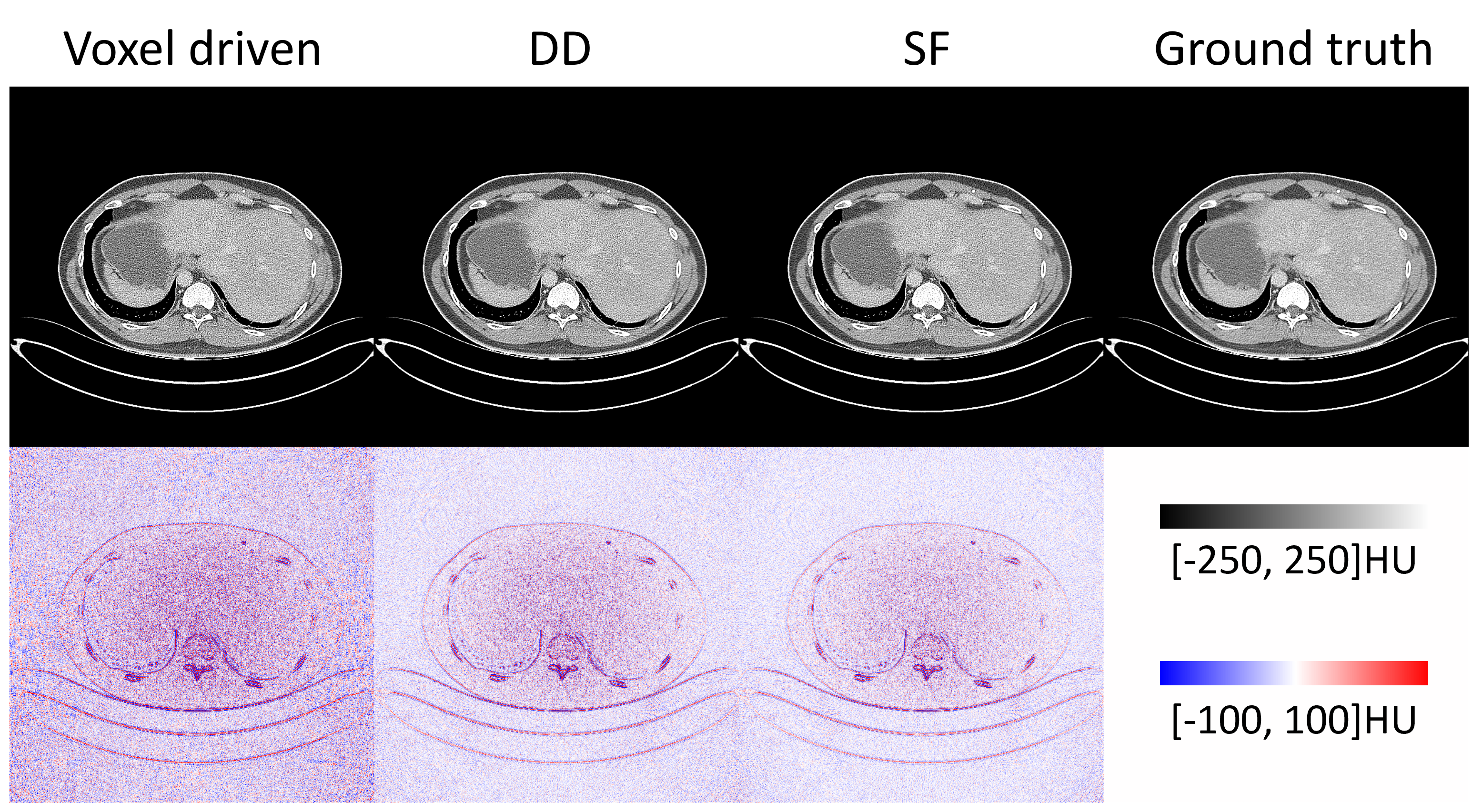}
        \caption{FBP reconstruction using the curved-detector projections from AAPM MAR challenge dataset. The bottom row display the difference images between CTorch reconstruction and the ground truth.}
        \label{fig:curve}
    \end{minipage}\hfill
    \begin{minipage}{0.48\textwidth}
        \centering
        \includegraphics[width=\textwidth]{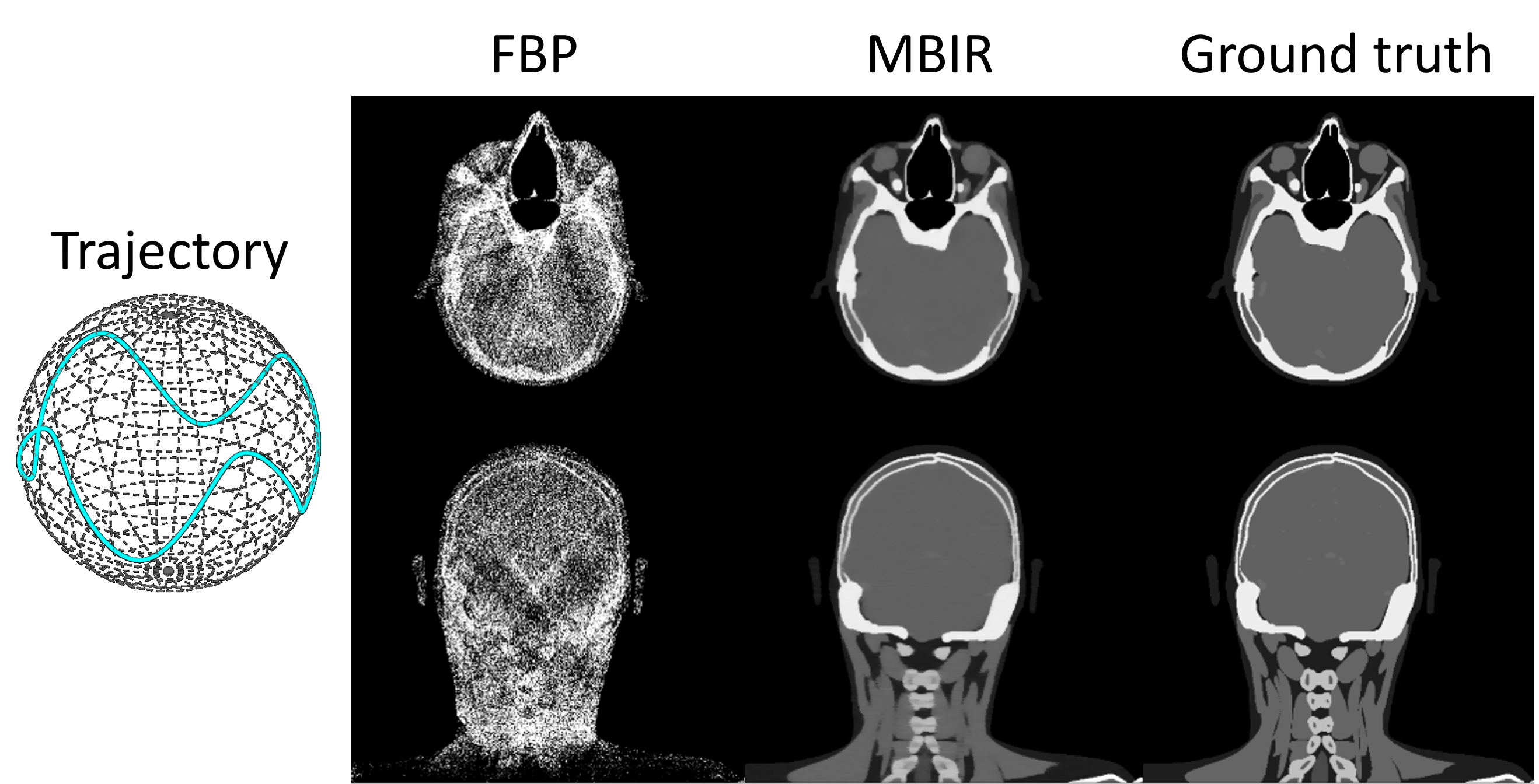}
        \caption{FBP and MBIR reconstruction of projections simulated with a non-circular scan trajectory.}
        \label{fig:mbir}
    \end{minipage}
\end{figure}

Next, we validate the CTorch projector for a curved detector using public sinogram data from the AAPM MAR challenge\cite{AAPM_CT_MAR_Challenge}. This dataset simulates 1D projections acquired using a curved detector with a 1.25-pixel lateral offset. The reconstruction results are summarized in Fig.~\ref{fig:curve}. The reconstructed images obtained with different projectors closely match the ground truth. The corresponding error maps further illustrate that no significant geometric distortions are present in the reconstruction, with errors primarily concentrated around sharp edges. These errors may be attributed to differences in the applied filtering function.

Finally, we validate the effectiveness of auto-differentiation through MBIR. As illustrated in Fig.~\ref{fig:mbir}, 200 head\cite{segars20104d} cone-beam projections are simulated with a sinusoidal scan trajectory. The objective function of MBIR is formulated as:
\begin{equation}
    \textbf{x}^*=\text{argmin}_\textbf{x} \|\textbf{Ax}-\textbf{y}\|_2^2 + \lambda\|\textbf{x}\|_{TV}
\end{equation}
The objective function is minimized with 200 iterations of the Adam optimizer where the gradient is computed via auto-differentiation (as shown in the Sec.\ref{sec:example}). Due to the irregular projection sampling pattern, FBP reconstruction exhibits substantial non-uniformity. In contrast, MBIR effectively improves the reconstruction accuracy, as evidenced by a clear depiction of bone and soft tissue boundaries. These results validate the effectiveness of auto-differentiation in conjunction with the CTorch projector.

\subsection{Computational Efficiency}
\begin{figure}[h]
    \centering
    \includegraphics[width=0.9\linewidth]{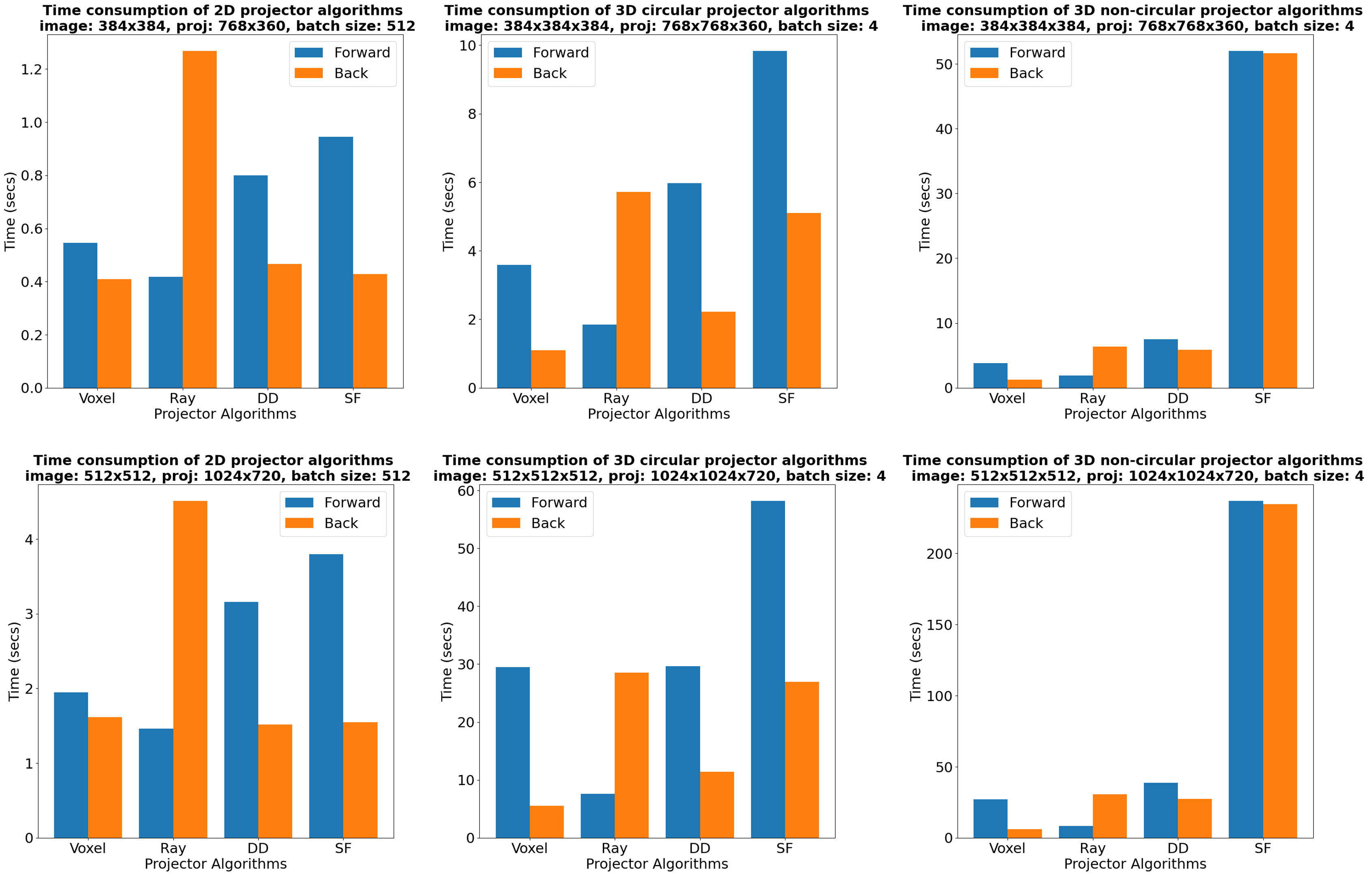}
    \caption{Computational efficiency of CTorch projectors for different geometries and different object sizes.}
    \label{fig:efficiency}
\end{figure}
We evaluated the efficiency of the CTorch projector on a workstation equipped with an AMD Ryzen 9 5950X CPU and an NVIDIA GeForce RTX 4090 GPU. The computation time for different geometries and object sizes is presented in Fig.~\ref{fig:efficiency}. It is important to note that the voxel-driven forward projector and ray-driven back projector are often not used in CT reconstruction imaging but are implemented in CTorch to provide gradients for their respective adjoint operators. In general, the voxel-driven projector offers the highest backprojection speed, while the ray-driven projector is the fastest for forward projection. The DD projector, which is considered more accurate than both voxel-driven and ray-driven projectors, exhibits slightly lower efficiency for paired forward and back projection. The SF projector, which theoretically provides the highest accuracy, has the lowest efficiency, particularly for non-circular geometries. 

Most current deep learning-based CT reconstruction research focuses on 2D cases. In this scenario, the DD projector can process 512 images in parallel within $1.3$~seconds for a $384^2$~image size and $4.7$~seconds for a $512^2$~image size, achieving millisecond-level processing per image with high numerical accuracy. In contrast, 3D projection operations are significantly slower due to the larger voxel count and the increased complexity of geometry-associated computations. The DD projector processes a $384^3$~volume in approximately $2$~seconds per volume. Current 3D deep learning-based reconstruction is typically applied to relatively small volume sizes ($\leq256^3$) due to GPU memory constraints, and the CTorch projector can achieve sub-second processing times for such volume sizes.

Regarding memory consumption, we found that most memory usage is allocated to storing the volume and projection data. As a result, memory consumption shows minimal differences between different projectors. Additionally, since data on the GPU is shared between the C++ and Python programs, no extra memory is required for data exchange.

\section{Conclusion}
We present CTorch, a novel GPU-accelerated, auto-differentiable projector toolbox designed for CT imaging with PyTorch compatibility. CTorch supports flexible scanner geometries, varied detector configurations, multiple projector algorithm options, and CUDA acceleration, while integrating with PyTorch auto-differentiation framework. These features enable broad applications in the field of CT imaging, including accurate CT simulations, efficient iterative reconstruction, and advanced deep-learning-based CT reconstruction.

\section*{Acknowledgments}
The authors wish to acknowledge Altea Lorenzon, Peiqing Teng, Yijie Yuan, Suyu Liao, Yiqun Ma, Mitchell Pelline, Shudong Li, Junyuan Li, Zimo Liu, Miao Qi, Yue Fan, and Donghyeon Lee for their assistance with code testing and debugging.

%Bibliography
\bibliographystyle{unsrt}  
\bibliography{references}

\end{document}